\documentclass[preprint,11pt]{article}

\usepackage{bm}
\usepackage{epsfig}
\usepackage{color}
\usepackage{amssymb}
\usepackage{amsmath}

\newcommand {\lab}[1]{\label{eq:#1}}

\newcommand {\be}[1]{\begin{equation}{\lab{#1}}}
\newcommand {\ee}{\end{equation}}
\newcommand {\bea}{\begin{eqnarray}}
\newcommand {\eea}{\end{eqnarray}}

\begin{document}

\title{Fermi-Pasta-Ulam model with long-range interactions: Dynamics and thermostatistics}

\author{
\textbf{Helen Christodoulidi$^{1,2}$, Constantino Tsallis$^{3,4}$ and Tassos Bountis$^{1}$}\\
$^{1}$Department of Mathematics, Division of Applied Analysis  
and \\ Center for Research and Applications of Nonlinear 
Systems (CRANS),\\ University of Patras, GR-26500 Patras, Greece. \\
$^2$Department of Mathematics, La Trobe University, \\
 Bundoora, VIC 3083, Australia\\
$^3$Centro Brasileiro de Pesquisas Fisicas and \\
National Institute of Science and Technology for Complex Systems,\\ 
Rua Xavier Sigaud 150, 22290-180 Rio de Janeiro-RJ, Brazil\\
$^4$Santa Fe Institute, 1399 Hyde Park Road, Santa Fe, NM 87501, USA}


\maketitle

\begin{abstract}
We introduce and numerically study a long-range-interaction generalization of the one-dimensional Fermi-Pasta-Ulam (FPU) $\beta-$ model. The standard quartic interaction is generalized through a coupling constant that decays as $1/r^\alpha$ ($\alpha \ge 0$)(with strength characterized by $b>0$). In the $\alpha \to\infty$ limit we recover the original FPU model. Through classical molecular dynamics computations we show that (i) For $\alpha \geq 1$ the maximal Lyapunov exponent remains finite and positive for increasing number of oscillators  $N$ (thus yielding ergodicity), whereas, for $0 \le \alpha <1$, it asymptotically decreases as $N^{- \kappa(\alpha)}$ (consistent with violation of ergodicity); (ii) The distribution of time-averaged velocities is Maxwellian for $\alpha$ large enough, whereas it is well approached by a $q$-Gaussian, with the index $q(\alpha)$ monotonically decreasing from about 1.5 to 1 (Gaussian) when $\alpha$ increases from zero to close to one. For $\alpha$ small enough, the whole picture is consistent with a crossover at time $t_c$ from $q$-statistics to Boltzmann-Gibbs (BG) thermostatistics. More precisely, we construct a ``phase diagram'' for the system in which this crossover occurs through a frontier of the form $1/N \propto b^\delta /t_c^\gamma$ with $\gamma >0$ and $\delta >0$, in such a way that the $q=1$ ($q>1$) behavior dominates in the $\lim_{N \to\infty} \lim_{t \to\infty}$ ordering ($\lim_{t \to\infty} \lim_{N \to\infty}$ ordering). 
\end{abstract}

                             
\newpage 
 
More than one century ago, in his historical book {\it Elementary Principles in Statistical Mechanics} \cite{Gibbs1902}, J. W. Gibbs remarked that systems involving long-range interactions will be intractable within his and Boltzmann's theory, due to the divergence of the partition function. This is of course the reason why no standard temperature-dependent thermostatistical quantities (e.g. specific heat) can possibly be calculated for the free hydrogen atom, for instance. Indeed, unless a box surrounds the atom, an infinite number of excited energy levels accumulate at the ionization value, which yields a divergent canonical partition function at any finite temperature.
 
In the present paper, we investigate the deep consequences of Gibbs' remark by focusing on the influence of the range of the interactions within an illustrative isolated classical system, namely a generalization of the Fermi-Pasta-Ulam (FPU) $\beta-$model \cite{FPU,Florence,Padova,Bountisetal,Tempestaetal}. Let us consider the Hamiltonian 
\begin{eqnarray}\label{ham}
{\cal H}=\frac{1}{2}\sum_{n=1}^{N} p_n^2 + \frac{1}{2}\sum_{n=0}^N (x_{n+1}-x_n)^2 
+ \frac{b}{4\widetilde N} \sum_{n=0}^{N} \sum_{m=n+1}^{N+1} \frac{(x_n-x_m)^4}{(m-n)^\alpha}=U(N) \;\;\;(b>0; \,\alpha \ge 0) \,,
\end{eqnarray}
with fixed boundary conditions (FBC), i.e. $x_0=x_{N+1}=p_0=p_{N+1}=0$.
Without loss of generality we have considered unit masses, and unit nearest-neighbor coupling constant; $p_n$ and $x _n$ are canonical conjugate pairs. 
At the fundamental state, all oscillators are still at $x_n=0$.
The nonlinear part of the potential energy {\it per particle} varies with $N$ like
\begin{eqnarray}\label{GALI:2}
   {\widetilde N}(N,\alpha)  \equiv    \frac{1}{N} \sum_{i=0}^{N} \sum_{j=i+1}^{N+1}  \frac{1}{(j-i)^\alpha} =
\frac{1}{N} \sum_{i=0}^{N} \frac{N+1-i}{(i+1)^\alpha} ~.
\end{eqnarray}

We notice that ${\widetilde N}(N,0)\simeq N/2$, and ${\widetilde N}(\infty,\alpha) =\zeta(\alpha)$, where $\zeta(\alpha)$ is the Riemann zeta function. Let us also remark that the $\widetilde N$ scaling is introduced in the Hamiltonian so as to make the total kinetic and potential energy {\it extensive} (i.e. proportional to $N$) for all values of $\alpha$. We note that the above scaling ${\widetilde N}(N,\alpha)$ applies to lattices with fixed boundary conditions and is only slightly different from the analogous scaling found in \cite{AnteneodoTsallis1998,CirtoAssisTsallis2013} meant for periodic boundary conditions (PBC).

The two limits (i) $\alpha \rightarrow 0$ and (ii) $\alpha \rightarrow  \infty $ are particularly interesting since they correspond to the extremal cases where, (i) each particle interacts equally with all others independently of the distance between them and (ii) only interactions with nearest neighbors apply, recovering exactly the Hamiltonian of the FPU-$\beta$ model. 

\begin{figure}
\begin{center}
\includegraphics[width=0.45\linewidth]{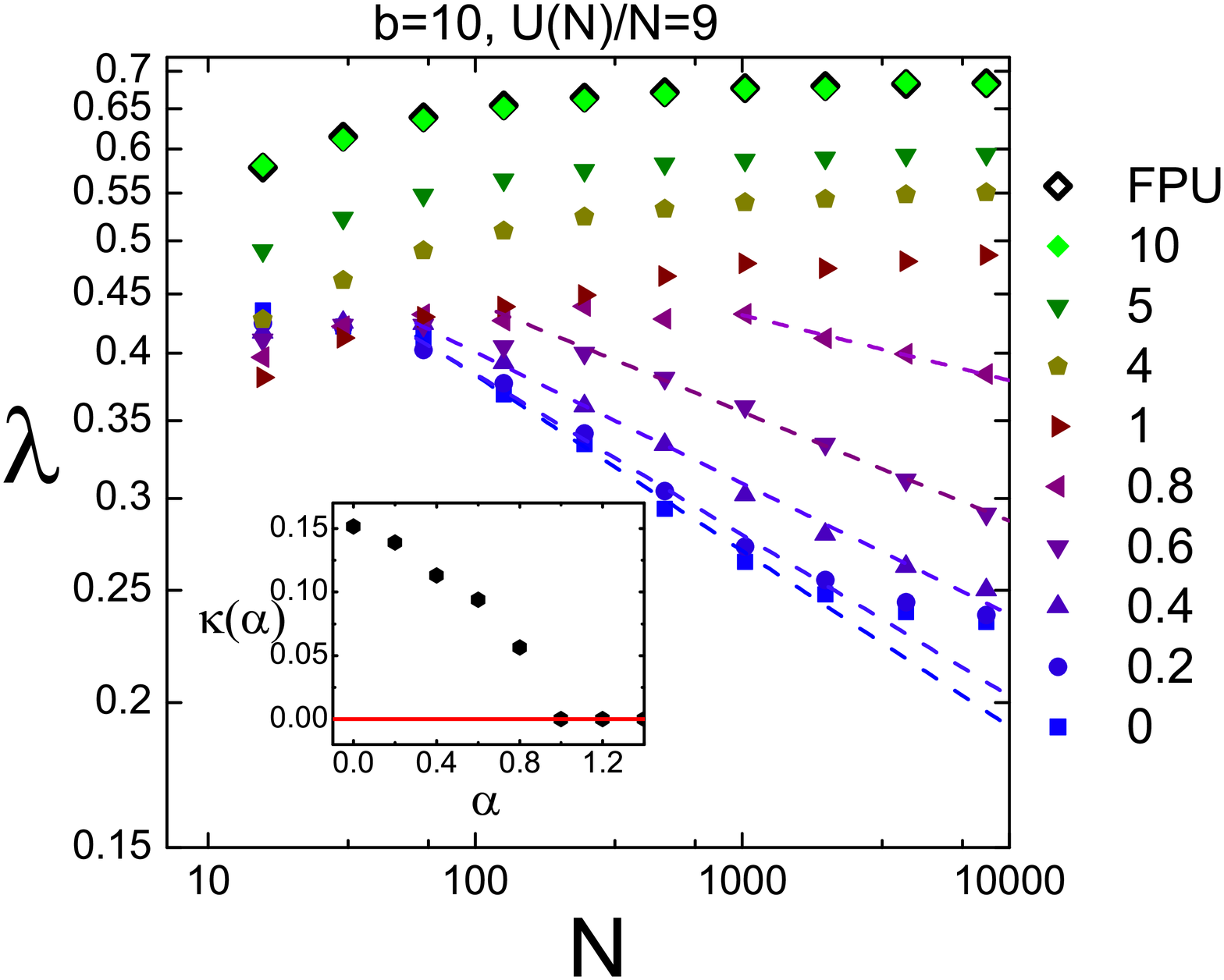}
\includegraphics[width=0.45\linewidth]{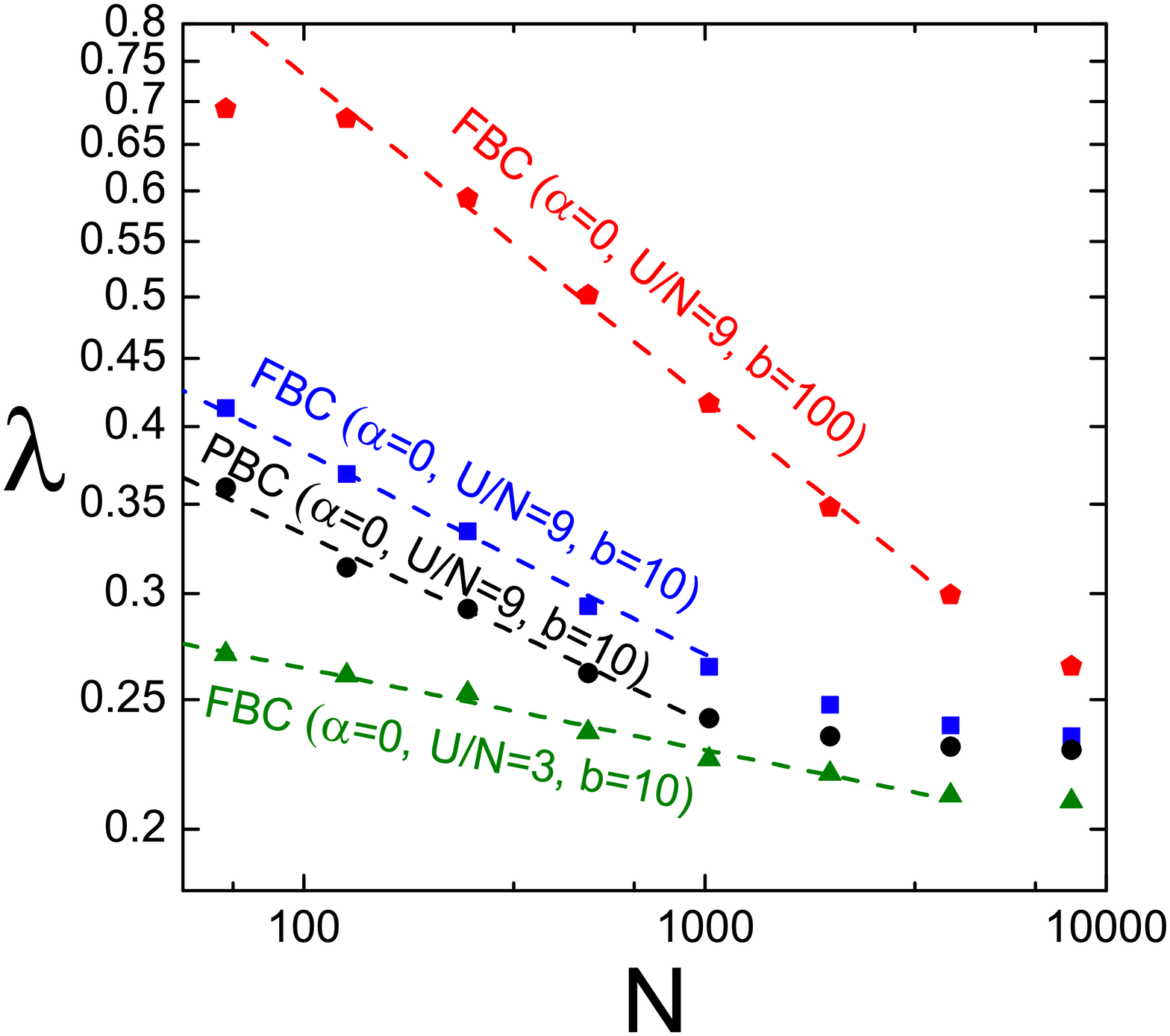}
\end{center}
\caption{  Maximal Lyapunov exponent for increasing $N$ calculated at $t=10^6$. Left panel: for various $\alpha $ values with $U(N)/N=9$, $b=10$ and FBC. Right panel: for various $U(N)/N$, $b$ values with $\alpha =0$ and both FBC, PBC.  \label{LEmax}}
\end{figure}

We note here that a significant difference of the present study from the generalized HMF model \cite{AnteneodoTsallis1998,PluchinoRapisardaTsallis2007,CirtoAssisTsallis2013} lies in the implementation of long range interactions {\it only in the quartic} part of the potential in (\ref{ham}) (the introduction of long-range interactions also in the quadratic term leads to similar results and will be addressed elsewhere). Our numerical results are obtained using the 4-th order Yoshida symplectic scheme with time--step such that the energy is conserved within 4 to 5 significant digits. The class of initial configurations we have chosen is of the ``water--bag'' type, i.e., zero positions and momenta drawn randomly from a uniform distribution.

Let us begin our study with a systematic investigation of the {\it largest Lyapunov exponent} $\lambda$ characterizing the ergodicity of the dynamics for different values of $\alpha $, $N$ and specific energies $u=U(N)/N$. In  Fig. \ref{LEmax} we have plotted  $\lambda$ versus the system size $N$ for different $\alpha $ values, ranging from 0 to 10. The critical value $\alpha =1$, similar to what was found in \cite{AnteneodoTsallis1998}, clearly distinguishes between the following two distinct regimes: 

(i) For $\alpha \geq 1$ the Lyapunov exponent $\lambda$ tends to stabilize at a finite and positive value as $N$ increases. 

(ii) For $\alpha < 1$ the largest Lyapunov exponents are observed to {\emph decrease} with system size as $N^{-\kappa(\alpha )}$, where the dependence of the exponent $\kappa(\alpha)$ on $\alpha $ is shown in the inset of Fig. \ref{LEmax}. 

We therefore expect that the system with short-range interactions tends to a BG type of equilibrium in the thermodynamic limit, characterized by ``strong'' chaos. On the other hand, the case of long-range interactions is  ``weakly'' chaotic.

\begin{figure}
\begin{center}
\includegraphics[width=0.45\linewidth]{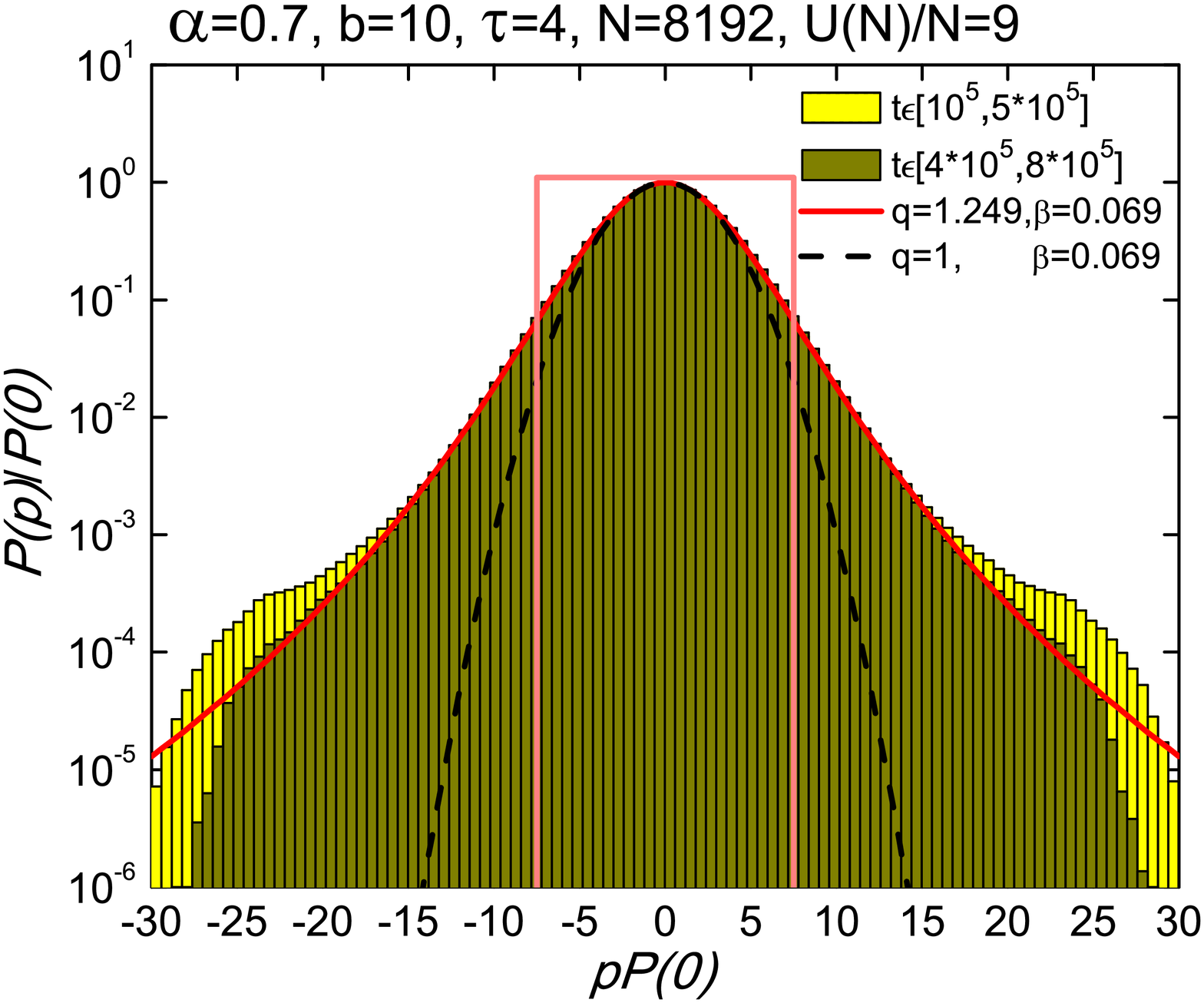} 
\includegraphics[width=0.45\linewidth]{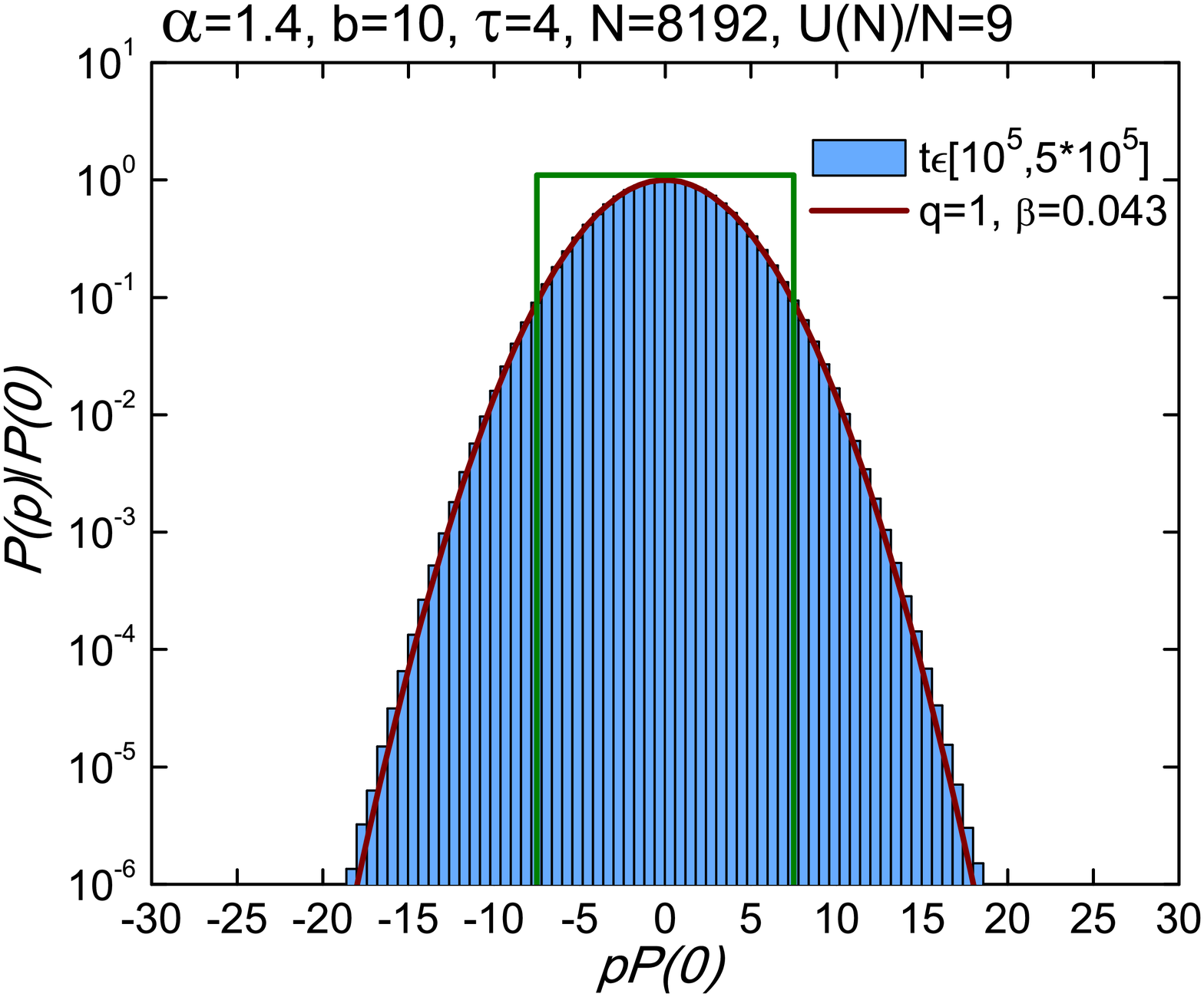} \\
\end{center}
\caption{  Time-averaged momentum distributions for the system with $N=8192$. Left panel: $\alpha =0.7$ for two different time intervals: The pdf seems to approach a $q$-Gaussian. Right panel: $\alpha =1.4$ and the distribution quickly approaches a Gaussian. 
\label{histogram}}
\end{figure}

\begin{figure}
\begin{center}
\includegraphics[width=0.5\linewidth]{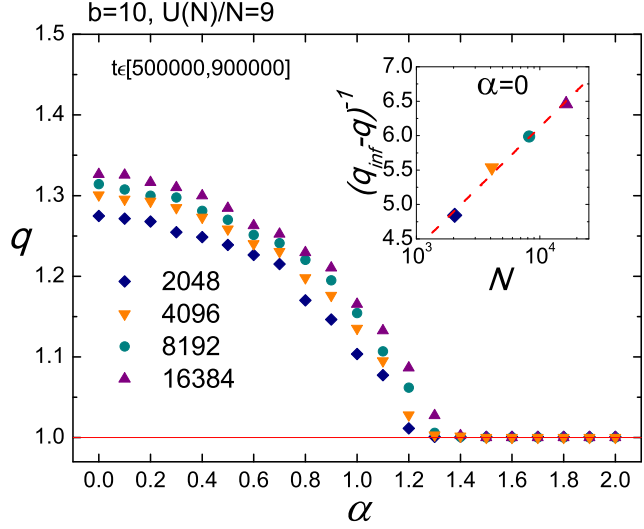} 
\end{center}
\caption{  $\alpha$-dependence of the index $q$ for $b=10$ and $U(N)/N=9$ averaged over 4 independent realizations when $N$ is 2048, 4096, 8192 and 2 realizations for $N=16384$, all taken in the time interval $t \in [5\cdot 10^5,9\cdot 10^5]$. {\it Inset:} $(q_{\infty }-q)^{-1}$ versus $N$, for the data of the main figure with $\alpha =0$;  $q_{\infty } $ has a value estimated around $1.48$, and is the intercept of the linear dependence of $q$ on $1/ \log N$. The fitting line shown is $(q_{ \infty } -q)^{-1}=1.76\log N -0.9 $.     
}\label{qalpha}
\end{figure}

\begin{figure}
\begin{center}
\includegraphics[width=0.5\linewidth]{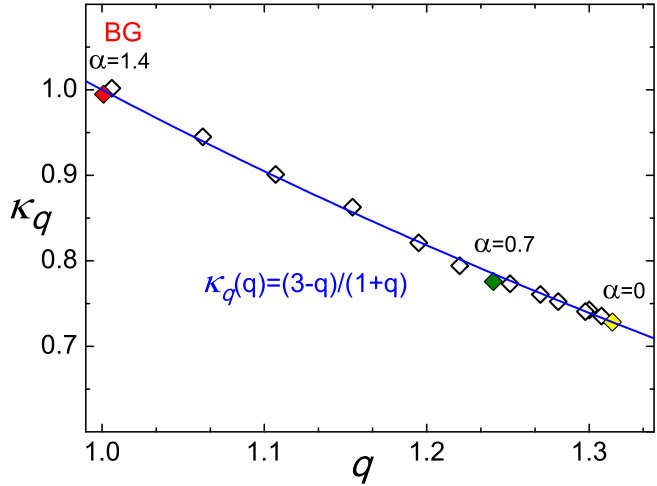} 
\end{center}
\caption{  $q$-dependence of the $q$-kurtosis $\kappa_q$ for typical values of $\alpha$, together with the analytical prediction $\kappa_q = (3-q)/(1+q)$ (blue curve) for the data of Fig.~\ref{qalpha}
which corresponds to $N=8192$. 
}
\label{qkurtosis}
\end{figure}

In order to check some of the expectations along the lines of nonextensive statistical mechanics (based on the nonadditive $q$-entropy)\cite{Tsallis1988,GellMannTsallis2004,Tsallis2009,Tsallis2014} and of the $q$-generalized Central Limit Theorem \cite{UmarovTsallisSteinberg2008}, we implement a molecular-dynamical computation of momentum distributions resulting from time averages of a single water--bag type initial condition of (\ref{ham}), calculated over the interval $[t_{min},t_{max}]$, where $t_{min}$ is such that the kinetic temperature $T\equiv 2 K(t)/N$ ($K(t)$ being the total kinetic energy of the system) stabilizes to a nearly constant value.

In particular, for each of the histograms of Fig. \ref{histogram}, we assign to each $p_i$ 
the number of times that the momenta fall in the $i$-th band, calculated repeatedly
for integer multiples of time (i.e. every $\tau =1$ for $N=2048$, $\tau =2$ for $N=4096$, $\tau =4$ for $N=8192$ etc. so that we always compare the same amount of data). 
Fig. \ref{histogram} displays the momentum distributions for $\alpha = 0.7$ and $1.4$ for $N=8192$. In the left panel two histograms are shown, one for the time interval $[10^5,5\cdot 10^5]$ and one for $[4\cdot 10^5,8\cdot 10^5]$, which are well fitted by the $q$--Gaussian pdf: 
\begin{equation}
P(p)=P(0) [1+ \beta(q-1)(pP(0))^2]^{1/(1-q)} , q \geq 1~,
\label{qgaussian}
\end{equation}
with $q=1.249$. This value of $q$ is nearly constant until $t=1.8 \cdot 10^6$.  For longer times $q$ is observed to decrease as a power law in time and tends to the value 1, which explains why we call this a quasi--stationary state (QSS) \cite{ABB2011}. In the Right panel of  Fig. \ref{histogram} on the other hand the distribution follows from the beginning a pure Gaussian pdf ($q\rightarrow1$ in (\ref{qgaussian})) with $\beta=0.043$.

The $q$--dependence on $\alpha $ is shown in Fig.~\ref{qalpha}, where the transition from 
$q$--statistics to BG--statistics is evident as $\alpha $ exceeds 1. Starting around $q \simeq 1.33$, $q$ reaches 1 at $\alpha =1.4$ for $N=16384$ particles calculated during the time interval $[5\cdot 10^5, 9\cdot 10^5]$. The data of Fig.~\ref{qalpha} is averaged over several realizations.

To check the robustness of our results with respect to $q$--statistcs, we have computed the $q$-generalized kurtosis (referred to as {\it $q$-kurtosis}
in \cite{TsallisPlastinoAlvarezEstrada2009, CirtoAssisTsallis2013}) defined as follows:
\begin{equation}
\kappa_q (q)=\frac{  \int_{-\infty}^\infty dp \,p^4 [P(p)]^{2q-1}  / \int_{-\infty}^\infty dp \, [P(p)]^{2q-1}  }{3 \Bigl[\int_{-\infty}^\infty dp \,p^2 [P(p)]^{q}  / \int_{-\infty}^\infty dp \, [P(p)]^{q}  \Bigr]^2 } \,.
\label{kurtosis}
\end{equation}
Using the $q$ values found in Fig.~\ref{qalpha} we plot in Fig.~\ref{qkurtosis} the numerical data of $q$-kurtosis vs. $q$ and find that it compares very well with the analytical curve $\kappa_q(q)=(3-q)/(1+q)$ obtained by substituting the $q$-Gaussian pdf (\ref{qgaussian}) in Eq. (\ref{kurtosis}).

\begin{figure} 
\begin{center}
\includegraphics[width=0.45\linewidth]{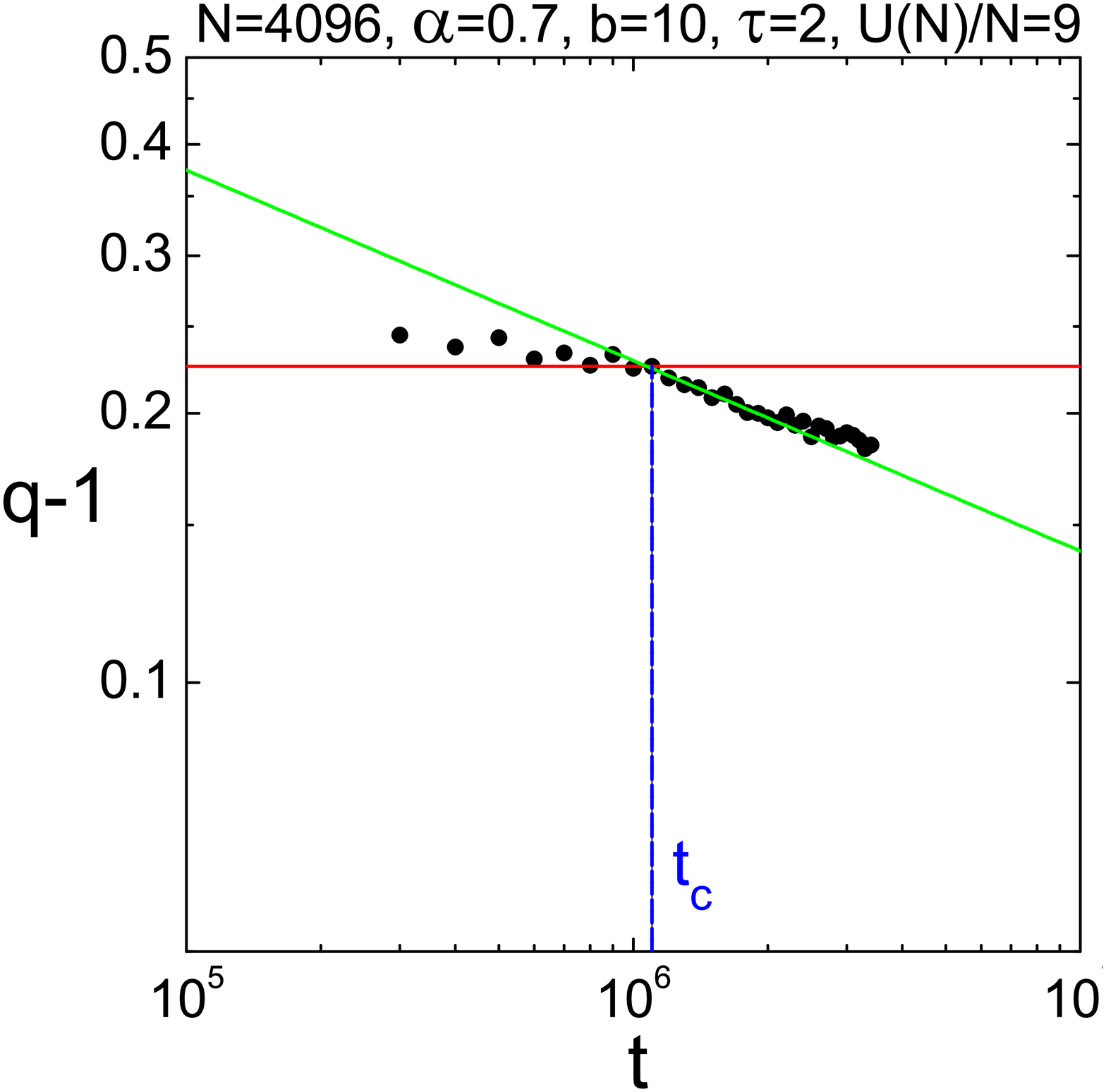}
\includegraphics[width=0.45\linewidth]{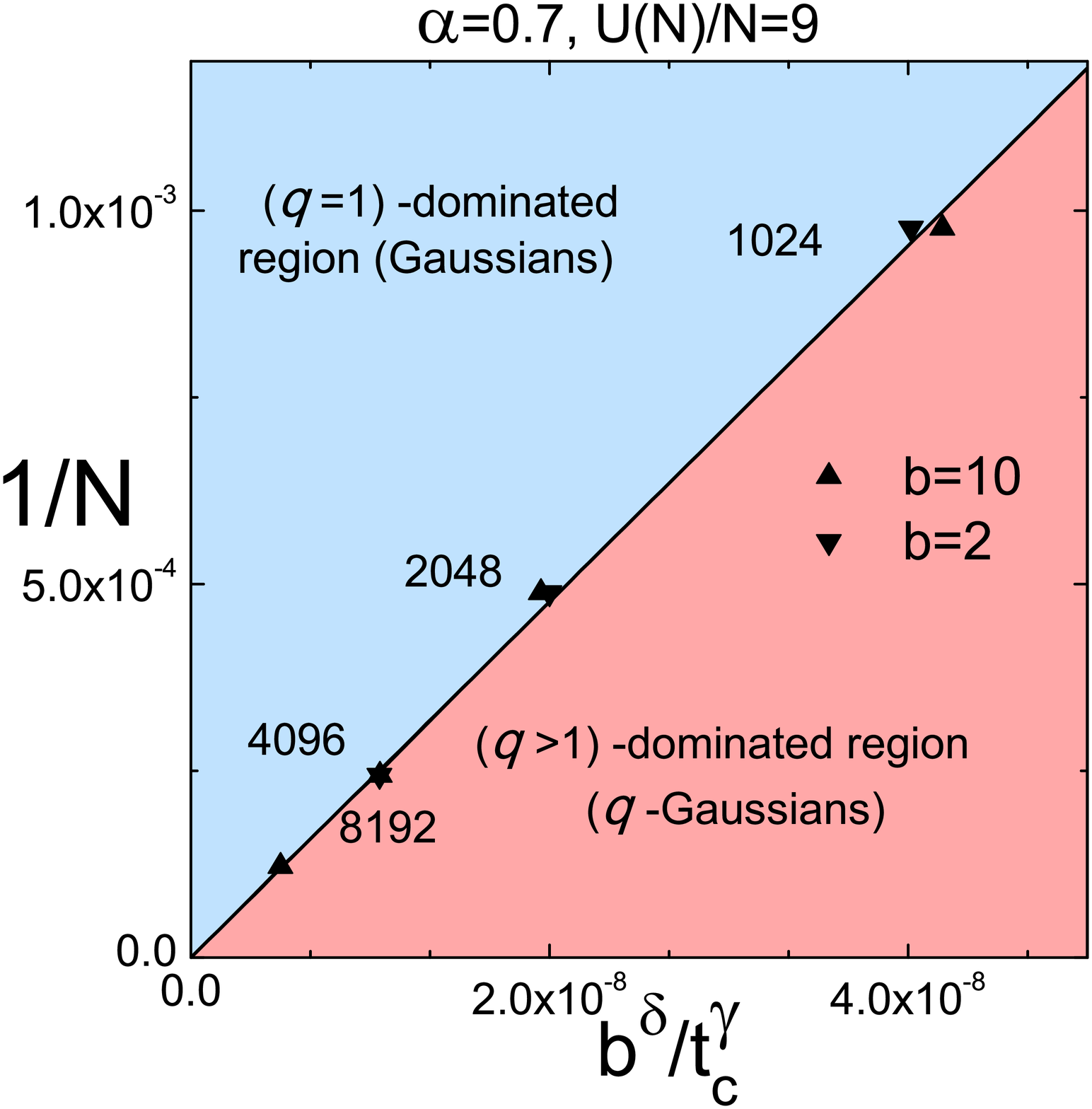}

\end{center}
\caption{  Left panel: Evolution of $q-1$ in double logarithmic scale also for $u \equiv U(N)/N=9$. Each point corresponds to 4 realizations of a time average in a running window of width $w=2\cdot 10^5$. $N$ is 4096 and the $t_c$ is defined as the intersection between the two lines. For fixed $N$, the QSS exists for times up to $t_c$, and the system slowly relaxes towards a BG behavior for times above $t_c$.
Right panel: Crossover frontier  between the Gaussian and $q$-Gaussian thermostatistical regions for the system sizes $N=1024, 2048, 4096, 8192$ at specific energy $u=9$, calculated for two cases, $b=2$ and $b=10$.  The fitting straight line is $1/N = D b^\delta/t_c^\gamma$, with $D =2.3818 \times 10^4$, $\delta=0.27048$, and $\gamma = 1.365$.} 
\label{crossover}
\end{figure}

Fig. \ref{crossover} (Right panel) displays the crossover between the two regimes in the form of a ``phase diagram'', in which, for fixed $b$, a straight line fit (in the $1/N$ vs. $1/t_c^{\gamma}$ plane) 
of the data $N \propto t^{\gamma }$ separates the two ``phases''. Each point in the graph corresponds to a value of $t=t_c$ representing the maximum time that $q$ remains constant, after which $q$ tends to the BG value $q=1$ following a power law (see Fig. \ref{crossover} Left panel). 

For high nonlinearity strength b = 10, the line separating the two domains
has a large slope.  When we decrease the nonlinearity to b = 2, the slope of
the boundary decreases.  In fact, the crossover frontier can be represented  for all $b$ by a single straight line given approximately by
\begin{equation}
\frac{1}{N}\sim D(\alpha,u) \frac{b^\delta}{t_c^\gamma}\, ~,
\end{equation}
where $D \ge 0$ depends on $\alpha$ (characterizing the range of the interactions) and on the energy per particle $u \equiv U(N)/N$, $\delta \simeq 0.27$ and $\gamma \simeq 1.36$. 

For $\alpha>1$, of course, $D$ vanishes and the system is expected to be uniformly ergodic, following BG statistics. For $\alpha<1$ on the other hand, all available numerical evidence strongly suggests that the system follows $q$-statistics during a non-ergodic QSS of ``weak chaos'', as if it were trapped (for large but finite $N$) in a subspace of the full phase space, where it lives for a very long time, until it eventually enters a ``strongly'' chaotic domain.

As a final summarizing remark, we emphasize the nonuniformity, for {\it long-range interactions} (i.e., $\alpha$ small enough), of the $(N,t) \to (\infty,\infty)$ limit implied by the diagram of Fig. \ref{crossover} (Right panel). Clearly, in the $\lim_{N \to\infty} \lim_{t \to\infty}$ ordering it is the $q=1$ behavior that prevails, while in the $\lim_{t \to\infty} \lim_{N \to\infty}$ ordering it is the $q>1$ statistics that becomes dominant. These results have been obtained from dynamical first principles (Newton's law), {\it without any a priori hypothesis} about entropy or whatever similar thermodynamical quantities.

\paragraph{Acknowledgments}
We acknowledge very fruitful remarks by L.J.L. Cirto. H.C. is grateful for the hospitality of La Trobe University, during October--December, 2013, where part of the work reported here was carried out. One of us (C.T.) gratefully acknowledges partial financial support by the Brazilian Agencies CNPq, Faperj and Capes. All of us acknowledge that this research has been co-financed  by the European Union (European Social Fund--ESF) and Greek national funds through the Operational Program `Education and Lifelong Learning' of the National Strategic Reference Framework (NSRF) - Research Funding Program: {\it Thales. Investing in knowledge society through the European Social Fund}.

\bibliographystyle{elsarticle-num}

\end{document}